\documentclass[12pt,preprint,url]{aastex}
\usepackage{amsbsy}
\usepackage{amsmath}
\newcommand{\be}{\begin{equation}}
\newcommand{\ee}{\end{equation}}

\def\ltsima{$\; \buildrel < \over \sim \;$}

\def\lsim{\lower.5ex\hbox{\ltsima}}
\def\gtsima{$\; \buildrel > \over \sim \;$}
\def\gsim{\lower.5ex\hbox{\gtsima}}

\shorttitle{Gravitational radiation from magnetar fallback}
\shortauthors{Melatos \& Priymak}

\begin{document}
\title{Gravitational radiation from magnetically funnelled
supernova fallback onto a magnetar}

\author{A. Melatos\altaffilmark{1}}
\email{amelatos@unimelb.edu.au}

\and

\author{M. Priymak\altaffilmark{1}}
\email{m.priymak@pgrad.unimelb.edu.au}

\altaffiltext{1}{School of Physics, University of Melbourne,
Parkville, VIC 3010, Australia}

\begin{abstract}
\noindent 
Protomagnetars spun up to millisecond rotation periods by supernova fallback
are predicted to radiate gravitational waves via hydrodynamic instabilities
for $\sim 10^2\,{\rm s}$ 
before possibly collapsing to form a black hole.
It is shown that magnetic funnelling of the accretion flow
(i) creates a magnetically confined polar mountain,
which boosts the gravitational wave signal,
and (ii) ``buries'' the magnetic dipole moment,
delaying the propeller phase and
assisting black hole formation.
\end{abstract}


\section{Introduction 
 \label{sec:fal1}}
Magnetars born rotating with millisecond periods have attracted
theoretical attention as the central engines powering
long-soft gamma-ray bursts
\citep{tho04,des08}
and optically brightened core-collapse supernovae
\citep{kas10}
and as sources of relativistic, Poynting-flux-dominated outflows
\citep{uso92,yi98,met11},
ultra-high-energy cosmic rays
\citep{aro03},
and gravitational waves
\citep{ste05,dal09,pir12}.
The latter signal is predicted to be detectable out to the Virgo Cluster 
by current-generation, long-baseline antennas 
like the Laser Interferometer Gravitational Wave Observatory (LIGO)
\citep{abb09}
at a rate of about one event per year.
Two gravitational radiation mechanisms have been analysed:
a permanent mass quadrupole created by magnetic stresses 
in the stellar interior
\citep{bon96,dal09,mas11},
and a transient quadrupole generated by hydrodynamic instabilities
like bar modes, which are excited when the protomagnetar spins up
during fallback
\citep{pir12}. 
In the latter context, it is crucial to understand how the rotation
and magnetization of the protomagnetar evolve, as fallback proceeds.
If the magnetic field is relatively high and/or the star rotates
relatively rapidly, the magnetic propeller effect
\citep{ill75}
shields the star from infalling material,
limiting the strength of the gravitational wave signal and
preventing the formation of a black hole
\citep{pir11}.

In this paper, we extend the magnetar fallback scenario by incorporating
magnetic funnelling of the accretion flow onto the magnetic poles
of the star
\citep{rom03}.
In other accreting systems, such as low-mass X-ray binaries,
magnetic funnelling strongly modifies the surface distributions
of mass and magnetic flux and hence observable properties like
the magnetic dipole moment
\citep{bro98,pay04},
equilibrium spin
\citep{wan11},
thermonuclear X-ray burst recurrence times, energies, and
harmonic content
\citep{pay06,mis10,cav11,pat12},
cyclotron lines
\citep{muk12},
and gravitational wave output
\citep{mel05,vig09a}.
Here we show that magnetar fallback is modified in two important ways:
(i) the accreting material forms a magnetically supported polar mountain,
whose gravitational radiation supplements the signal
from the internal magnetic quadrupole and hydrodynamic instabilities;
and (ii) polar magnetic burial reduces the magnetic dipole moment
and hence the effectiveness of the propeller mechanism,
making it easier to form a black hole.

The paper is structured as follows.
In \S\ref{sec:fal2}, 
we introduce an idealized, general-purpose model of magnetar fallback 
proposed recently
\citep{pir11,pir12}.
We investigate how magnetic funnelling modifies
gravitational wave emission, mass capture,
and black hole formation in the context of the model 
in \S\ref{sec:fal3} and \S\ref{sec:fal4} respectively.
Conclusions and a survey of the limitations of the calculation
are presented in \S\ref{sec:fal5}.
We emphasize at the outset that the results pertain specifically
to the scenario,
addressed by the authors above and many others,
where a magnetar is born spinning fast,
with a rotation period of a few milliseconds.
At least some magnetars may not be born in this manner.
X-ray observations of three supernova remnants associated with
anomalous X-ray pulsars and soft gamma-ray repeaters in the Milky Way imply
that the explosion energy of their progenitors is 
close to the canonical supernova value of $10^{51}\,{\rm erg}$, 
arguing against fast rotation at birth in three out of $\sim 20$
known objects
\citep{vin06}.
\footnote{
Strictly speaking, \citet{vin06} argued against fast rotation
lasting longer than a (brief) initial time window.
Strong gravitational radiation immediately after birth can nullify
the issue of over-powering the remnant
\citep{dal09}; cf. \citet{lai01}.
}

\section{Protomagnetar evolution during fallback
 \label{sec:fal2}}
We begin by summarizing the key ingredients of
magnetar fallback.
In \S\ref{sec:fal2a}, we adopt the standard, parametrized
prescriptions for the propeller torque, accretion rate,
and hydrodynamic instability threshold favoured in the literature
\citep{pir11,pir12}.
In \S\ref{sec:fal2b}, we calculate how the mass ellipticity $\epsilon$
and magnetic dipole moment $\mu$ evolve as functions of the 
accreted mass $M_{\rm a}$, applying the rigorous theory of
polar magnetic burial developed originally for recycled neutron stars
\citep{pay04,pri11}.
At every step we emphasize the idealizations in the model,
which are unavoidable.
The messy physics of fallback, especially its geometry,
is incompletely understood even for $\sim 10^{12}\,{\rm G}$ fields,
let alone for magnetars 
\citep{ber10}.
A related calculation without magnetic funelling was performed by 
\citet{wat02}
in the context of gravitational radiation from r-modes during
supernova fallback.
In addition to bar-mode instabilities and the propeller effect,
\citet{wat02} treated viscous damping
and the star's thermal response to accretion and r-mode heating,
effects which are neglected here.

\subsection{Accreted mass and angular velocity
 \label{sec:fal2a}}
Accretion during fallback has been studied thoroughly
in the context of collapsars,
``mild'' core-collapse events in which part of the stellar mantle
initially explodes then stalls and implodes.
Numerical simulations of collapsars which treat 
the hydrodynamics, shock physics, and neutrino transport 
in detail find accretion rates in the range $10^{-4}$ to 
$10^{-2}\, M_{\odot}\,{\rm s^{-1}}$ 
lasting for $10^3$ to $10^4\,{\rm s}$
\citep{mac01,zha08}.
Fallback passes through early and late stages,
with 
$\dot{M}_{\rm early} \approx 10^{-3} \eta t^{1/2}
 M_{\odot} \, {\rm s^{-1}}$ and
$\dot{M}_{\rm late} \approx 50 t^{-5/3}
 M_{\odot} \, {\rm s^{-1}}$ respectively,
where $t$ is the time after core bounce (in s),
and $\eta$ depends sensitively on the explosion energy, 
$E_{\rm s}$,
with $0.1 \leq \eta \leq 10$ for 
$0.3 \leq E_{\rm s} / 10^{51}\,{\rm erg} \leq 1.2$.
The transition occurs at $t\sim 10^2\,{\rm s}$,
after the oxygen shell falls in.
We adopt the convenient parametrization
$\dot{M}_{\rm a}
 = (\dot{M}_{\rm early}^{-1} + \dot{M}_{\rm late}^{-1})^{-1}$
for the total accretion rate,
$\dot{M}_{\rm a}$,
introduced by \citet{pir11}.
For $\eta$ in the above range, $\dot{M}_{\rm a}$ is mostly
high enough to overcome the outward ram pressure of the magnetar's
neutrino- and Poynting-flux-driven wind,
even when the system is spherically symmetric
\citep{pir11,pir12}.

The angular velocity of the magnetar, $\Omega(t)$,
evolves in response to three torques,
$N_{\rm dip}$, $N_{\rm a}$, and $N_{\rm gw}$,
defined below,
according to
$I\dot{\Omega} = N_{\rm dip} + N_{\rm a} + N_{\rm gw}$,
where 
$I=0.35MR^2$ is the moment of inertia in terms of the
stellar mass $M$ and radius $R$
\citep{lat01}.
Figure 6 of \citet{lat01} gives 
$0.2 \leq I/MR^2 \leq 0.5$
for a range of plausible equations of state;
we adopt a central value here.
The magnetic dipole torque $N_{\rm dip}$ obeys
the standard vacuum formula,
$N_{\rm dip} = - \mu^2 \Omega^3 / (6 c^3)$,
as long as the light cylinder (radius $c/\Omega$) lies inside 
the hydromagnetic lever arm,
given by the nominal Alfv\'{e}n radius
$r_{\rm m} = \mu^{4/7} (GM)^{-1/7} \dot{M}_{\rm a}^{-2/7}$.
For millisecond magnetars, this is a fair approximation,
setting aside extended-dipole corrections
\citep{mel97}.
The accretion torque $N_{\rm a}$ takes either sign,
depending on the ratio of its mechanical and magnetic components,
and is controlled by the fastness parameter
$\omega=(r_{\rm m}/r_{\rm c})^{3/2}$,
where $r_{\rm c} = (GM / \Omega^2)^{1/3}$
is the corotation radius
\citep{gho79}.
Here we follow \citet{pir11} and set
$N_{\rm a} = (1-\omega) (GMr_{\rm m})^{1/2} \dot{M}_{\rm a}$.
A more careful treatment of the disk-magnetosphere boundary
and hence $N_{\rm a}$,
to include episodic accretion in a corotation-limited disk
\citep{rap04,dan11}
and radiation pressure
\citep{and05},
\footnote{
Radiation pressure effects are modified by the
strong polarization dependence of the relevant opacities
at magnetar field strengths,
a problem currently under investigation
\citep{van13}.
}
lies outside the scope of this paper;
our aim here is to explore how magnetic funnelling affects
magnetar fallback irrespective of the specific mode of disk accretion.
The gravitational wave torque $N_{\rm gw}$ is treated phenomenologically,
following \citet{pir11},
by setting $N_{\rm gw}= - N_{\rm a} \theta(\beta-\beta_{\rm c})$,
where $\theta(\dots)$ is the Heaviside step function,
$\beta(M,R,\Omega)$ is the
ratio of the star's kinetic and gravitational potential energies,
and $\beta_{\rm c}$ is the threshold for 
radiation-reaction-driven ($\beta_{\rm c}=0.14$), 
viscosity-driven (0.14), 
and bar mode (0.27) instabilities.
Qualitatively, gravitational wave friction acts
to cancel the other torques,
once a suitable instability is triggered,
with $N_{\rm gw} \approx - N_{\rm a} - N_{\rm dip}$ and
$|N_{\rm dip}| \ll |N_{\rm a}|$ typically.
In \S\ref{sec:fal2b}, we add to $N_{\rm gw}$ the secular contribution
from the quadrupole associated with a polar magnetic mountain.

The mass of the star, $M(t)$, evolves according to 
$\dot{M} = \dot{M}_{\rm a} \theta(1-\omega)$,
where $\dot{M}_{\rm a}$ is the parametrized fallback rate above,
and the propeller effect shuts off accretion
for $\omega > 1$.
This oversimplifies things:
even under propeller conditions, 
some of the infalling plasma penetrates to the stellar surface
due to instabilities at the disk-magnetosphere boundary
\citep{rap04,rom05,dan11}.
We bundle the uncertainties surrounding propeller leakage specifically
and disk dynamics generally into the parameter $\eta$,
an approach justified approximately elsewhere through
one-zone, Shakura-Sunyaev disk calculations
\citep{pir11,pir12}.
To stay consistent with previous work, 
the initial mass is set to $M(0)=1.4 M_{\odot}$,
the distinction between gravitational and baryonic mass is ignored,
and $R$ is held constant as $M$ increases,
a fair approximation for most equations of state until just before
black hole formation
\citep{lat01}.

\subsection{Polar magnetic burial
 \label{sec:fal2b}}
The strong magnetic field of a magnetar funnels the fallback
accretion flow preferentially onto the magnetic poles to form
a mountain.
Funnelling is imperfect,
due to Rayleigh-Taylor mixing on the boundary of the polar flux tube
and filamentation caused by obliquity,
but simulations confirm that it occurs for a wide range
of parameters and geometries
\citep{rom03,kul08}.
As matter splashes down onto the poles,
it slides sideways under its own weight,
dragging along magnetic field lines by flux freezing.
Thus the magnetic field is compressed at the equator
and pushes back to confine the mountain,
while the radial component of the polar field and hence $\mu$
are reduced
\citep{mel01,pay04}.

To calculate accurately the magnetic field structure ${\bf B}({\bf x})$ 
and density profile $\rho({\bf x})$ of a polar mountain
is a subtle task.
Order-of-magnitude pressure balance arguments underestimate the
mass quadrupole $\sim 10^4$-fold by implicitly imposing outflow
boundary conditions at the edge of the polar cap and hence underestimating
the compression of, and tension in, the equatorial field, e.g.\
\citet{bro98} and references therein.
To avoid this,
one must solve the force balance equation
(quasistatic equilibrium; Alfv\'{e}n crossing time-scale
$\ll M_{\rm a} / \dot{M}_{\rm a}$)
simultaneously with a mass-flux constraint equation enforcing
flux freezing and equatorial magnetic compression.
Writing 
${\bf B} = (r\sin\theta)^{-1} \nabla\psi(r,\theta)
 \times \nabla\phi$
in spherical polar coordinates $(r,\theta,\phi)$,
where $\psi(r,\theta)$ is a scalar flux function,
we obtain
\begin{equation}
 \Delta^2\psi
 =
 - \frac{dF(\psi)}{d\psi}
 \left\{
  1 -
  \frac{(\Gamma-1) (\phi-\phi_0)} 
   {\Gamma K^{1/\Gamma} [F(\psi)]^{(\Gamma-1)/\Gamma}}
 \right\}^{1/(\Gamma-1)}
\label{eq:fal1}
\end{equation}
and
\begin{equation}
 F(\psi)
 =
 \frac{K}{(2\pi)^\Gamma}
 \left( \frac{dM}{d\psi} \right)^\Gamma
 \left[
  \int_C ds\, r\sin\theta | \nabla\psi |^{-1}
  \left\{ \dots \right\}^{1/(\Gamma-1)}
 \right]^{-\Gamma}
\label{eq:fal2}
\end{equation}
for the force balance equation and mass-flux constraint respectively
for a polytropic equation of state $P=K\rho^\Gamma$,
where $P$ is the pressure,
$\phi(r)$ is the gravitational potential in the Cowling approximation,
$\phi_0$ is the gravitational potential at the stellar surface
before accretion begins,
$\Delta^2$ is the Grad-Shafranov operator, 
the integral in (\ref{eq:fal2}) is computed along the field line
$\psi={\rm constant}$,
and the braces $\{ \dots \}$ in (\ref{eq:fal2})
contain the same expression as the braces in (\ref{eq:fal1}).
The reader is referred to the literature for a detailed derivation
of (\ref{eq:fal1}) and (\ref{eq:fal2}) and a key to the notation
\citep{pay04,pri11,muk13a}.
A vital point concerns the barometric function $F(\psi)$ 
and mass-flux ratio $dM/d\psi$,
i.e.\ the mass (initial plus accreted) per unit flux
enclosed within $(\psi,\psi+d\psi)$.
Some authors guess $F(\psi)$ in (\ref{eq:fal1}) and drop (\ref{eq:fal2})
for convenience
\citep{bro98,mel01,muk13a},
but in general such a guess produces the wrong amount of equatorial
magnetic compression.
In reality
$F(\psi)$ is determined uniquely by accretion
through the relevant initial value problem and $dM/d\psi$.
In equilibrium, this path-specific information is stored in (\ref{eq:fal2}),
which maps one-to-one the initial (pre-accretion) and final
states to preserve flux freezing at all intermediate steps, 
as required by
the mass continuity and magnetic induction equations of 
ideal magnetohydrodynamics (MHD)
\citep{pri11}.

\begin{figure}
\begin{center}
\includegraphics[width=12cm,angle=0]{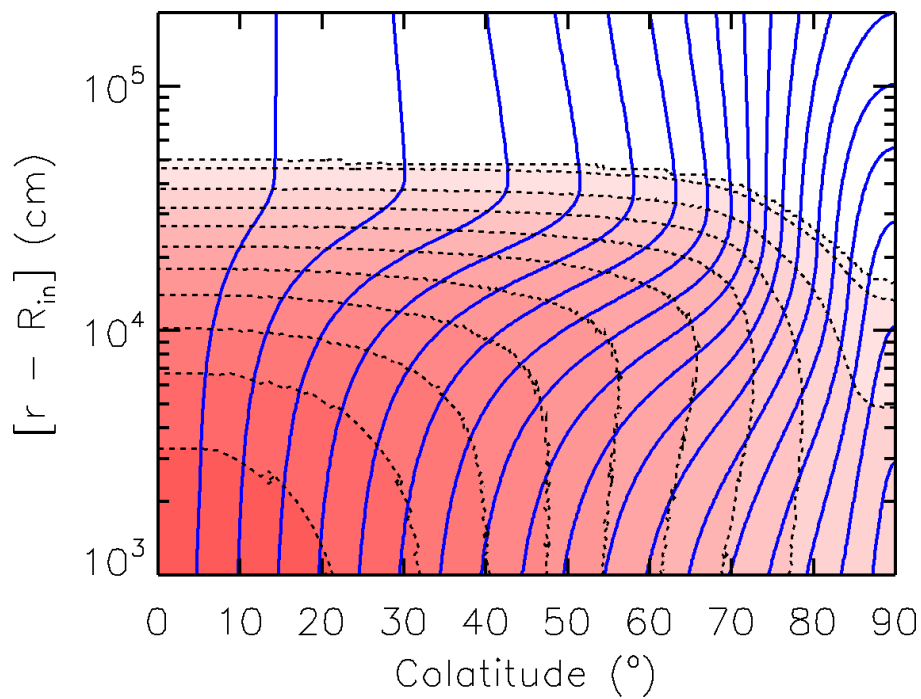}
\end{center}
\caption{
Hydromagnetic structure of a polar magnetic mountain on a magnetar,
showing magnetic field lines (solid blue curves) and isodensity
contours (dashed black curves and greyscale in red).
One quadrant of a meridional section is displayed,
with altitude above the base (in ${\rm cm}$)
on the vertical axis
and colatitude  on the horizontal axis.
Mountain parameters:
$M_{\rm a} = 1.2\times 10^{-3} M_\odot = 1.6 M_{\rm c}$,
$\mu_{\rm i} = 5\times 10^{32} \, {\rm G\,cm^3}$,
equation of state model D from \citet{pri11},
$\mu=0.6\mu_{\rm i}$,
and $\epsilon=5\times 10^{-4}$.
}
\label{fig:fal1}
\end{figure}

Figure \ref{fig:fal1} displays the equilibrium structure of a 
typical magnetic mountain 
$\sim 10\eta^{-1/2} \, {\rm s}$ after fallback begins,
with initial magnetic dipole moment
$\mu_{\rm i} = 5\times 10^{32} \, {\rm G\,cm^3}$,
accreted mass $M_{\rm a} = 1.2\times 10^{-3} M_\odot$,
and line-tying boundary conditions
[see \S{2.1} and \S{2.2} in \citet{pri11} for details].
From the solid contours, one observes that the magnetic field
is compressed into an equatorial band, 
with $B_r \ll B_\theta$ near the pole and
$B_\theta \ll B_r$ on either side of the equator.
The magnetic tension, directed along the field-line radius of curvature, 
confines the accreted matter into a mound,
whose isodensity surfaces are drawn as dashed contours.
The equilibrium is calculated for $M_{\rm a} = 1.6 M_{\rm c}$,
where 
$M_{\rm c} = 
 3\times 10^{-3} (\mu_{\rm i} / 10^{33} \, {\rm G\,cm^3})^2 M_\odot$
is the characteristic burial mass for a polytropic equation of state
with nondegenerate neutrons
[model D in \citet{pri11}].
Other polytropes with $1 \leq \Gamma \leq 5/3$
give qualitatively similar results
[see Figure 8 in \citet{pri11}].
By constructing a quasistatic sequence of equilibria with
increasing $M_{\rm a}$,
one finds that the magnetic dipole and mass quadrupole moments
scale roughly as
$\mu = \mu_{\rm i} (1+M_{\rm a}/M_{\rm c})^{-1}$
and
$\epsilon = (M_{\rm a}/M_\odot) (1+M_{\rm a}/M_{\rm c})^{-1}$
respectively
\citep{shi89,pay04,zha06,pri11};
polar magnetic burial screens $\mu$,
even as the mountain builds up,
and $|{\bf B}|$ rises at the equator.
In Figure \ref{fig:fal1}, for example, we have
$\mu=0.6 \mu_{\rm i}$ and $\epsilon = 5\times 10^{-4}$
for $M_{\rm a} = 1.6 M_{\rm c}$,
within $50\%$ of the predictions from the rule-of-thumb formulas.
The formula for $\epsilon$, 
which implies $\epsilon \leq M_{\rm c}/M_\odot$,
is deliberately conservative,
to avoid overpredicting the resulting gravitational wave signal.
In reality,
$\epsilon$ rises gradually above $M_{\rm c} / M_\odot$
for $M_{\rm a} \gtrsim M_{\rm c}$
in time-dependent MHD simulations that grow the mountain from scratch
[see Figure 5 in \citet{vig09a}],
accompanied by transient, localized, loss-of-equilibrium events 
(cf.\ instabilities),
where the magnetic field pinches off at isolated points 
to form topologically disconnected loops without disrupting
the main body of the mountain
\citep{kli89,pay04,vig08,muk13a}.

It is natural to wonder whether the stressed magnetic configuration
in Figure \ref{fig:fal1} is stable on the Alfv\'{e}n time-scale
$\tau_{\rm A}$.
The answer, counterintuitively, seems to be yes.
Independent MHD simulations by two groups using the solvers ZEUS
\citep{pay07,vig08}
and PLUTO
\citep{muk13a,muk13b}
confirm stability in two and three dimensions.
A self-consistent solution of (\ref{eq:fal1}) and (\ref{eq:fal2}),
when imported into ZEUS or PLUTO,
initially experiences the undular submode of the Parker instability,
but it is not disrupted;
after releasing a ``magnetic blister'', 
the mountain settles down to a new equilibrium with
$\epsilon$ reduced by $50$ to $70\%$ 
and mass loss $\lesssim 1\%$
\citep{vig08}.
Magnetic line tying at the inner boundary stabilizes the undular
submode and switches off the interchange submode and growing modes
in the continuous spectrum completely
\citep{vig08, vig09c}.
\citet{muk13a} found unstable, pressure-driven, filamentary,
toroidal modes at the periphery of filled and hollow mountains, but
(i) these modes do not disrupt the mountain overall,
consistent with \citet{vig08};
(ii) they are suppressed when fixed-gradient boundary conditions
at the edge of the polar cap are replaced by north-south symmetry 
at the equator, 
to allow properly for stabilization by the
equatorial magnetic belt, cf.\ \citet{lit01};
and (iii) they arise from equilibria satisfying (\ref{eq:fal1})
but not (\ref{eq:fal2}) (see above).
Grid-refinement tests buttress these findings
\citep{vig08,muk13a}.
For example, the linear growth rate of the undular submode
is observed to vary $\propto \, ({\rm grid \,\, scale})^{-1/2}$
and is grid-limited, as theory predicts,
but the final, saturation values of $\mu$ and $\epsilon$ are found
to be independent of the grid scale
\citep{vig08}.

A magnetic mountain does not relax resistively or by sinking
on protomagnetar time-scales ($\lesssim 10^4\,{\rm s}$)
\citep{vig09b,wet10},
nor is it disrupted by resistive instabilities.
The buried polar magnetic field and hence $\mu$ resurrect 
on the shorter of the ohmic diffusion and Hall time-scales
$\tau_{\rm d}$ and $\tau_{\rm H}$,
with 
${\rm min}(\tau_{\rm d},\tau_{\rm H}) \gtrsim 1\,{\rm yr}$
even with enhanced accretion-driven heating
\citep{vig09b,vig13}.
Likewise,
subduction and meridional redistribution
occur long after the magnetar spins down and stops emitting 
gravitational waves at a detectable level, if no black hole forms
\citep{cho02}.
The values of $\mu$ and $\epsilon$ calculated from
(\ref{eq:fal1}) and (\ref{eq:fal2}) are within a factor of two
of ZEUS mountains grown from scratch on a soft surface,
as long as the boundary at $r=R_{\rm in}$
in Figure \ref{fig:fal1} is set deep enough,
so that the mountain is light compared to the substrate,
its base does not move much laterally,
and magnetic line tying stays a good approximation
\citep{wet10}.
In MHD simulations where the electrical resistivity is boosted
artificially,
the mountain is stable on the tearing mode time-scale,
$(\tau_{\rm d} \tau_{\rm A})^{1/2} \sim 10^4\,{\rm s}$
\citep{vig08}.

\section{Gravitational radiation
 \label{sec:fal3}}
A magnetic mountain supplements the gravitational radiation
from instabilities.
We find in \S\ref{sec:fal3a}
that the added signal is substantial.
The improvement in detectability is potentially even greater,
as shown in \S\ref{sec:fal3b},
because the mountain radiates longer than instabilities under certain conditions,
and its waveform is cleaner
(cf.\ nonlinear fluid motions near centrifugal break-up).
We estimate the gravitational wave strain assuming that
the neutron star survives then look at how the conclusions
are modified by black hole formation in \S\ref{sec:fal4}.

\subsection{Peak wave strain
 \label{sec:fal3a}}
\begin{figure}
\begin{center}
\includegraphics[width=9cm,angle=0]{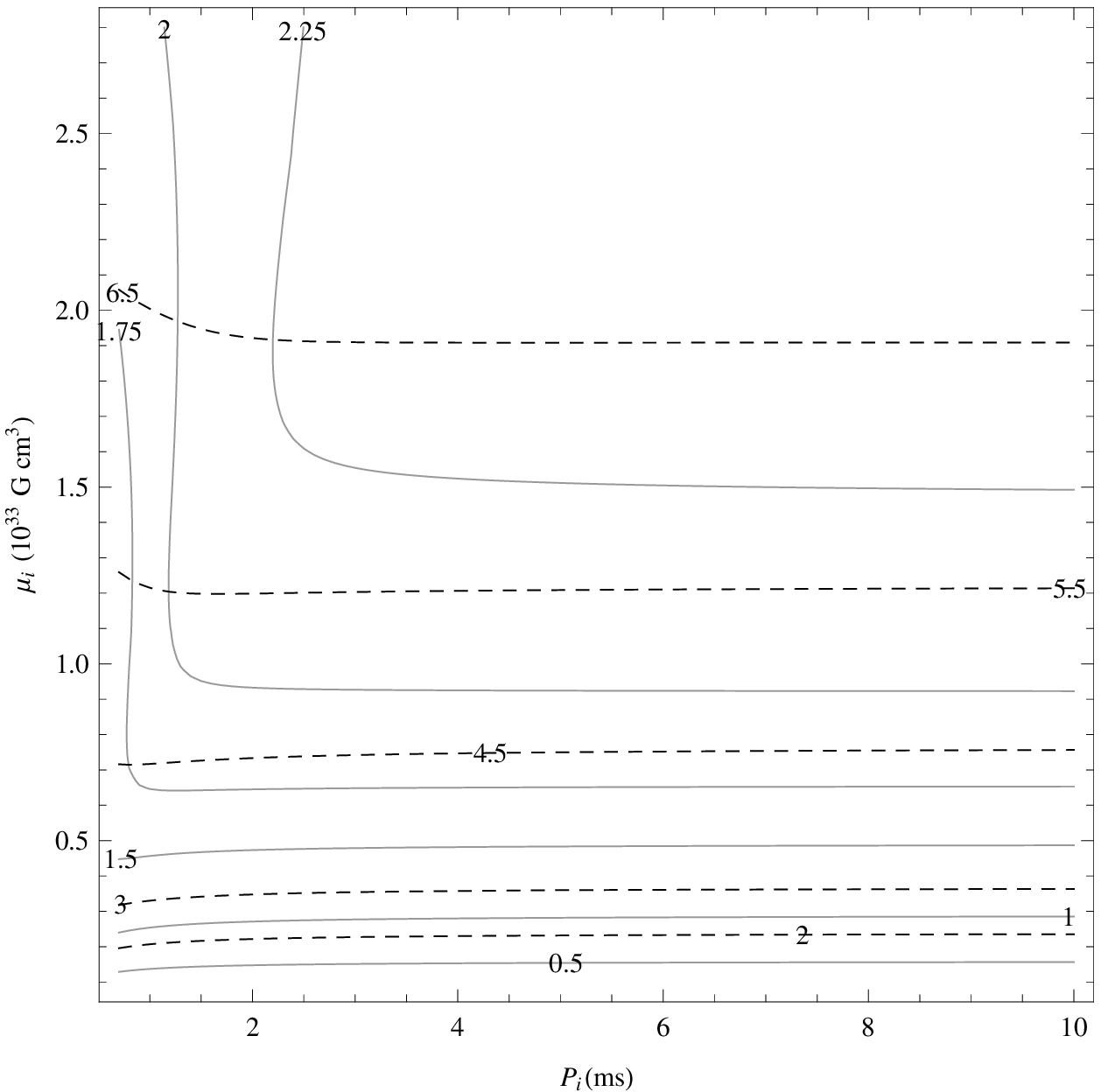}
\includegraphics[width=9cm,angle=0]{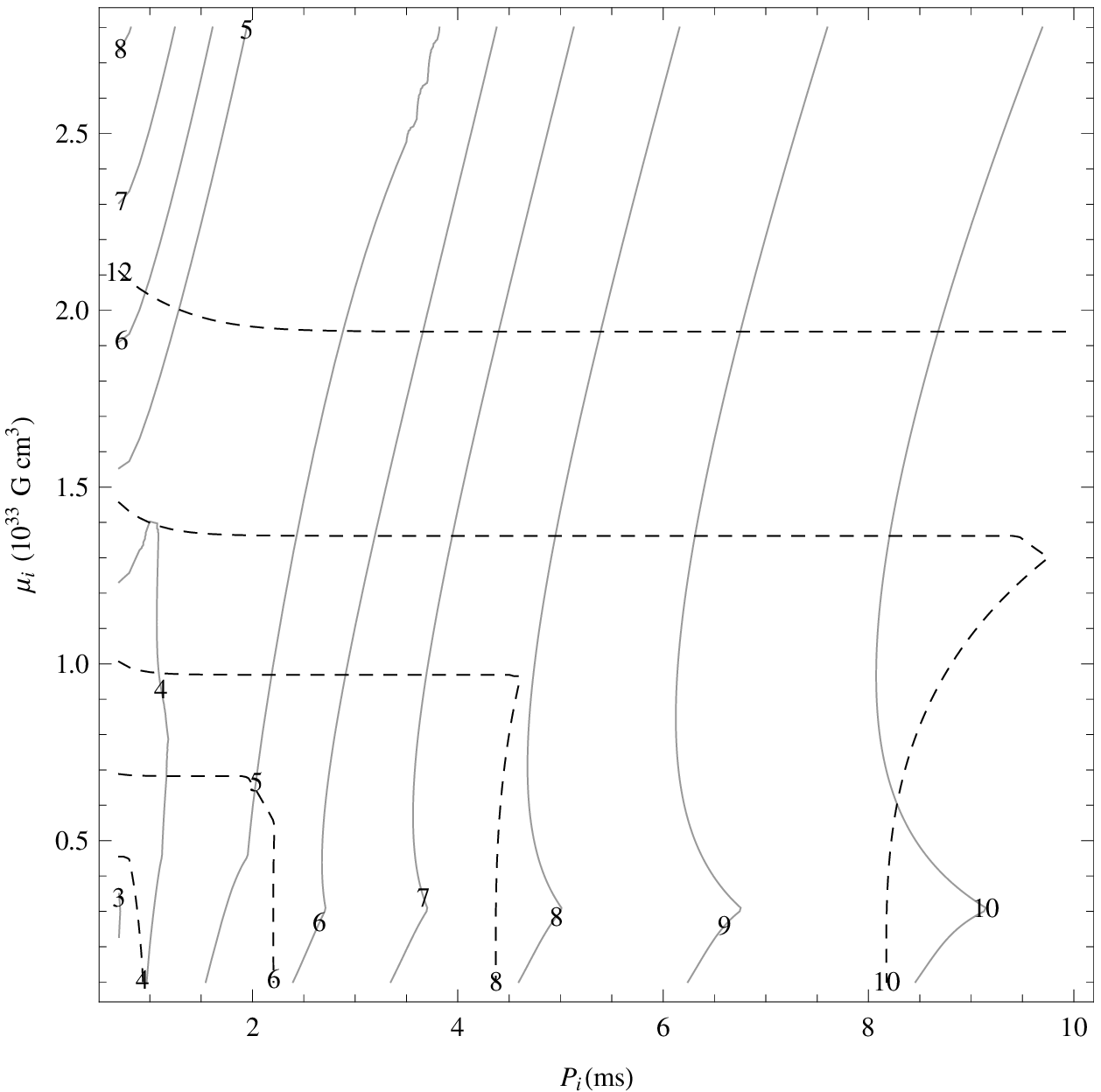}
\end{center}
\caption{
Contours of peak gravitational wave strain 
$h_{\rm max} = {\rm max}[h_0(t)]$
(in units of $10^{-23}$)
as a function of initial spin period $P_{\rm i}$ 
(in units of ${\rm ms}$) and
initial magnetic dipole moment $\mu_{\rm i}$
(in units of $10^{33}\,{\rm G\,cm^3}$),
excluding and including magnetic dipole reduction by burial
(solid and dashed curves respectively).
({\em Top panel.})
Radiation from magnetic mountain only.
({\em Bottom panel.})
Radiation from magnetic mountain and hydrodynamic instabilities.
Parameters:
$\eta=1$, 
$\beta_{\rm c}=0.14$,
$M_{\rm c} = 3\times 10^{-3} 
 ( \mu_{\rm i} / 10^{33}\,{\rm G\,cm^3})^2$,
$M(0)=1.4M_\odot$, $R=10\,{\rm km}$,
$D=1\,{\rm Mpc}$.
}
\label{fig:fal2}
\end{figure}

\begin{figure}
\begin{center}
\includegraphics[width=11cm,angle=0]{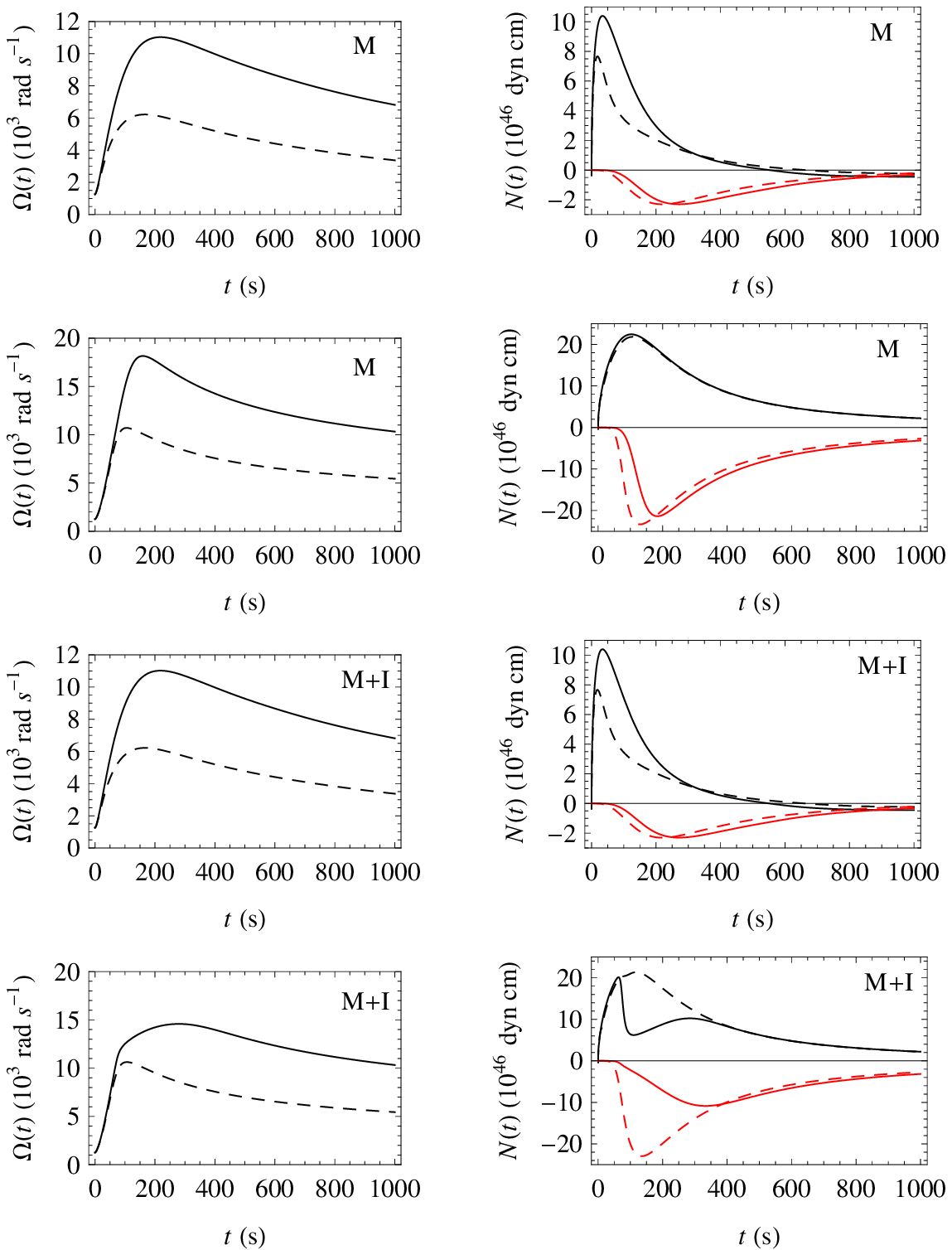}
\end{center}
\caption{
Rotational evolution of a protomagnetar experiencing fallback accretion.
({\em Left column.}) Angular velocity $\Omega$ versus time $t$.
({\em Right column.}) Torques due to accretion ($N_{\rm a}$; black curves)
and gravitational radiation ($N_{\rm gw}$; red curves) versus $t$.
Four scenarios are considered,
according to whether or not a magnetic mountain forms and radiates,
hydrodynamic instabilities are excited and radiate,
and $\mu$ is reduced by burial.
({\em Top row.})
Mountain forms, $\mu$ constant, instabilities switched off.
({\em Second row.})
Mountain forms, $\mu$ reduced, instabilities switched off.
({\em Third row.})
Mountain forms, $\mu$ constant, instabilities switched on.
({\em Bottom row.})
Mountain forms, $\mu$ reduced, instabilities switched on.
In every panel, the solid and dashed curves correspond to 
low magnetization
($\mu_{\rm i}=0.7\times 10^{33}\,{\rm G\,cm^3}$,
$P_{\rm i} = 5\,{\rm ms}$)
and high magnetization
($\mu_{\rm i}=1.5\times 10^{33}\,{\rm G\,cm^3}$,
$P_{\rm i} = 5\,{\rm ms}$)
respectively.
Other parameters as for Figure \ref{fig:fal2}.
The plotted quantities vary by $\sim 15\%$ over the range
$0.2 \leq I/MR^2 \leq 0.5$.
}
\label{fig:fal5}
\end{figure}

Figure \ref{fig:fal2}
displays contours of the peak gravitational wave strain 
$h_{\rm max} = {\rm max} [ h_0(t) ]$
as a function of $\mu_{\rm i}$ and the initial spin period $P_{\rm i}$.
The maximum is computed for $0\leq t \leq 10^4\,{\rm s}$
and typically occurs at 
$t\approx t_{\rm pk} = 85 \eta^{-6/13} \, {\rm s}$,
i.e.\ the accretion time-scale,
which does not depend on $\mu_{\rm i}$ or $P_{\rm i}$.
To clarify the physics, we examine separately
the consequences of magnetic funnelling
and hydrodynamic instabilities in the top and bottom panels
of Figure \ref{fig:fal2}.
To help the reader interpret Figure \ref{fig:fal2},
we also plot representative examples of the star's rotational evolution
and dominant torque components 
($N_{\rm a}$, $N_{\rm gw} \gg N_{\rm dip}$)
in Figure \ref{fig:fal5}
for four scenarios (top to bottom rows),
defined by whether or not magnetic funnelling occurs,
hydrodynamic instabilities are switched on,
and $\mu_{\rm i}$ is low (solid curves) or high (dashed curves).
The rotational evolution differs between the scenarios,
as we now discuss.

In the top panel,
we switch off artificially the instabilities,
so that they contribute neither to $h_0$ nor to $N_{\rm gw}$,
and focus on the radiation from the mountain,
with characteristic strain
$h_0(t) = 2 G \Omega(t)^2 \epsilon(t) I(t)/ (c^4 D)$,
where $D$ is the distance to the source.
The solid contours describe what happens, 
when the mountain quadrupole grows with $t$,
yet $\mu$ stays constant ---
an unlikely scenario, 
given the physics of magnetic burial in \S\ref{sec:fal2b}, 
but still instructive as a stepping stone to the full problem.
For $\mu_{\rm i} \gtrsim 0.7 \times 10^{33} \, {\rm G\,cm^3}$,
the star quickly attains magnetocentrifugal equilibrium
[$\omega\approx 1$, 
$\Omega \approx 7\times 10^3 
 (\mu / 10^{33}\,{\rm G\,cm^3})^{-6/7} \, {\rm rad\,s^{-1}}$]
at
$t\approx t_{\rm pk}$,
getting there either by spinning up by accretion
($P_{\rm i} \gtrsim  1\,{\rm ms}$)
or spinning down by the propeller effect
($P_{\rm i} \lesssim  1\,{\rm ms}$).
It then spins down gradually over $\sim 10^3\,{\rm s}$
mainly under the action of $N_{\rm gw}$,
with $N_{\rm a}$ assisting the deceleration 
for $t\gtrsim 5\times 10^2\,{\rm s}$,
and $N_{\rm dip}$ contributing $\sim 10\%$ of the total torque
(see Figure \ref{fig:fal5}, top row, right panel).
The instability threshold is never crossed;
$\beta$ peaks at $\approx 0.05$,
so it is consistent to switch off the instabilties in this regime.
The peak wave strain increases gently with $\mu_{\rm i}$
according to
$\epsilon \propto M_{\rm c} \propto \mu_{\rm i}^2$
($M_{\rm a} \gg M_{\rm c}$),
$\Omega \propto \mu_{\rm i}^{-6/7}$ 
(magnetocentrifugal equilibrium),
and hence 
$h_{\rm max} \propto \epsilon \Omega^2 \propto \mu_{\rm i}^{2/7}$.
The $h_{\rm max}$ scaling steepens at low $P_{\rm i}$,
where the solid contours turn upward,
because the propeller physics asserts itself earlier.

For $\mu_{\rm i} \lesssim 0.7 \times 10^{33} \, {\rm G\,cm^3}$
in the above scenario (Figure \ref{fig:fal5}, top row, dashed curves),
the star spins up towards its centrifugal limit,
\footnote{
Following \citet{pir11}, we do not model centrifugal break up.
Hence there are brief time intervals in some scenarios in Figure \ref{fig:fal5},
during which $\Omega$ approaches the centrifugal limit for low $\mu_{\rm i}$
(see solid curves in Figure \ref{fig:fal5}),
and the model breaks down.
In a real system, the instability back-reaction self-adjusts 
to prevent break up, and the results do not change qualitatively.
}
entering the propeller phase relatively late at
$t \gtrsim 5\times 10^2\,{\rm s} \gg t_{\rm pk}$.
It then spins down, but more gradually than for
$\mu_{\rm i} \gtrsim 0.7 \times 10^{33} \, {\rm G\,cm^3}$
because, by the time the star enters the propeller phase,
$N_{\rm a} \propto \dot{M}_{\rm a}$ is well below its peak
($N_{\rm dip} \sim N_{\rm a}$ here).
The gravitational wave signal is also weaker, 
as the solid contours indicate,
and as expected from $h_{\rm max} \propto \mu_{\rm i}^{2/7}$.
However, the predicted $h_{\rm max}$ is unrealistic,
because $\beta$ promptly exceeds $\beta_{\rm c}$ at $t\approx t_{\rm pk}$
in the low-$\mu_{\rm i}$ regime, 
i.e.\ the switched-off instabilities would switch on rapidly
in reality to emit gravitational radiation and 
spin down the star through $N_{\rm gw}$.
The latter behavior is discussed further below with reference
to the bottom panel of Figure \ref{fig:fal2}.

Now suppose that $\mu$ diminshes through magnetic burial,
while the mountain grows,
and the instabilities remain artificially switched off.
The results are described by the dashed contours in the top panel
of Figure \ref{fig:fal2};
see also the second row of Figure \ref{fig:fal5}.
The contours in Figure \ref{fig:fal2} are nearly horizontal.
Burial suppresses the propeller effect observed at low $P_{\rm i}$
and high $\mu_{\rm i}$ in the solid contours,
because at $t\gtrsim t_{\rm pk}$ we have $M_{\rm a} \gg M_{\rm c}$
and hence $\mu \ll \mu_{\rm i}$ everywhere in the plotted region.
Moreover, with $\mu \ll \mu_{\rm i}$ everywhere, 
$N_{\rm a}$ behaves similarly for low and high $\mu_{\rm i}$;
compare the dashed and solid black curves in the second row 
of Figure \ref{fig:fal5}.
The accretion torque rapidly spins up the star towards
torque balance ($N_{\rm a}+N_{\rm gw}\approx 0$),
while maintaining $\omega < 1$ for all $t$ 
except $t\lesssim 10\,{\rm s} \ll t_{\rm pk}$,
when $M_{\rm a}$ is still less than $M_{\rm c}$.
The wave strain $h_{\rm max} \propto \Omega^2$
is $\approx 4$ times greater than for the solid contours,
because $\Omega$ peaks at the stall frequency
\citep{bil98},
which turns out to be about double the magnetocentrifugal
equilibrium frequency
(see above).
The scaling for the stall frequency,
$\Omega \propto \epsilon^{-2/5} \propto \mu_{\rm i}^{-4/5}$,
implies
$h_{\rm max} \propto \epsilon \Omega^2 \propto \mu_{\rm i}^{2/5}$,
which is confirmed by inspecting the dashed contours.
The trend with $P_{\rm i}$ is weak,
as the propeller is inactive.
The stall frequency is reached at $t\approx t_{\rm pk}$,
after which $\dot{M}_{\rm a}$ and hence $N_{\rm a}$ drop away rapidly,
and $h_0(t)$ decays on the characteristic gravitational wave spin-down time-scale 
$I\Omega/N_{\rm gw}
 \propto \mu_{\rm i}^{-4/5}$,
evaluated at the stall point,
with
$I\Omega/N_{\rm gw} \sim 10^3\,{\rm s}$ typically
(noting $|N_{\rm dip}| \ll |N_{\rm gw}|$).
The decay time-scale influences the detectability of the signal,
as discussed in \S\ref{sec:fal3b}.
For $\mu_{\rm i} \gtrsim 1 \times 10^{33} \, {\rm G\,cm^3}$,
$\beta$ peaks below $\beta_{\rm c}$,
so switching off artificially the instabilities does not affect the conclusions. 
At lower $\mu_{\rm i}$, the conclusions are affected, as discussed below.

Finally, let us switch on the hydrodynamic instabilities.
The results including (dashed contours) and excluding (solid contours)
$\mu$ reduction by magnetic burial are presented in the 
bottom panel of figure \ref{fig:fal2}
and the third and fourth rows of Figure \ref{fig:fal5}.
The instability-sourced contribution to the wave strain
is given by
$h_0(t) = [5 G N_{\rm a}(t)]^{1/2} / [8c^3 D^2 \Omega(t)] ]^{1/2}$
from torque balance 
[$N_{\rm a} = N_{\rm gw}$;
see \S\ref{sec:fal2a} and
\citet{pir11}].
\footnote{
By summing the mountain and instability contributions to 
$h_0(t)$ and $N_{\rm gw}(t)$,
we assume implicitly that they are independent,
which is debatable;
e.g.\ hydrodynamic instabilities may move the magnetic footpoints 
at the base of the mountain and hence affect $\epsilon (t)$.
Likewise, once $\beta$ exceeds $\beta_{\rm c}$, 
it is unclear whether $N_{\rm a}$ self-adjusts
to balance all of $N_{\rm gw}$ or just its instability-sourced component.
Large-scale magnetohydrodynamic simulations outside the scope of this paper
are needed to resolve these subtle issues properly.
\label{foot:fal1}
}
We set the threshold at $\beta_{\rm c}=0.14$;
the results are qualitatively similar for $\beta_{\rm c}=0.27$.
Consider first the solid contours.
They are distorted at the left edge of the plot 
($P_{\rm i} \lesssim 2\,{\rm ms}$),
because the propeller effect spins down the star initially,
before $\Omega$ rises to a local maximum 
$\gtrsim 3\times 10^3\,{\rm rad\,s^{-1}}$
at $t\approx t_{\rm pk}$.
Along the lower edge of the plot,
for $\mu_{\rm i} \lesssim 0.3\times 10^{33} \, {\rm G\,cm^3}$
(where the contours display kinks),
accretion spins up the star towards its centrifugal limit,
triggering hydrodynamic instabilties;
$\beta$ stays above $\beta_{\rm c}$ for $t\gtrsim t_{\rm pk}$,
overshooting as far as $\beta \approx 0.2$,
and the star spins down electromagnetically on the time-scale
$\sim 10^3 (\mu_{\rm i} / 10^{33}\,{\rm G\,cm^3})^{-2} \,{\rm s}$, 
with $N_{\rm gw}$ balancing $N_{\rm a}\propto \dot{M}_{\rm a}$,
as the instabilities operate.
For $\mu_{\rm i} \gtrsim 0.3\times 10^{33} \, {\rm G\,cm^3}$,
no instabilities are triggered, 
$\beta < \beta_{\rm c}$ decreases monotonically with $t$,
and the rotational evolution in the first and third rows of 
Figure \ref{fig:fal5} is similar.

On the other hand,
the dashed contours in Figure \ref{fig:fal2} describe what happens,
when $\mu$ is reduced by burial.
Across the whole plot,
we find $\omega < 1$ for all $t$, 
i.e.\ there is no propeller effect.
Likewise $N_{\rm dip}$ is insignificant.
For $\mu_{\rm i} \gtrsim 1\times 10^{33} \, {\rm G\,cm^3}$,
no instabilities are triggered.
The star spins up to the stall frequency, satisfying 
$N_{\rm a}\approx N_{\rm gw}$
\citep{bil98},
then spins down in response to the mountain component of $N_{\rm gw}$
(with $|N_{\rm gw}| \approx 1.5 |N_{\rm a}|$)
on the time-scale
$\sim I\Omega/N_{\rm gw} \propto \mu_{\rm i}^{-4/5}$.
The latter evolution resembles the dashed contours in the top panel 
of Figure \ref{fig:fal2}
and the dashed curves in the second row of Figure \ref{fig:fal5},
as the mountain quadrupole is significant.
For
$\mu_{\rm i} \lesssim 1 \times 10^{33} \, {\rm G\,cm^3}$,
instabilities are triggered,
$N_{\rm a}$ drops sharply due to the instability back-reaction
(see solid curve at $t\approx 70\,{\rm s}$ in Figure \ref{fig:fal5},
fourth row, right panel),
and $\Omega\sim 10^4\,{\rm rad\,s^{-1}}$ decays slowly
under the action of $N_{\rm gw}$ and $N_{\rm dip}$, 
which is much reduced by burial.

\subsection{Signal-to-noise ratio
 \label{sec:fal3b}}
An accurate estimate of the signal-to-noise ratio (SNR) for a protomagnetar
undergoing fallback requires detailed calculations of the waveform
and Monte Carlo simulations of the search pipeline, both of which lie
outside the scope of this paper.
However, it is useful to convert the wave strain predictions
in Figure \ref{fig:fal2} into a rough detectability measure,
in order to clarify the relative importance 
of the peak strain, $h_{\rm max}$, and signal lifetime, $T_{\rm obs}$.
To give a flavor of what is possible,
we consider two extremes:
(i) a matched filter search,
which assumes optimistically that one can track the phase
of the signal coherently for its duration,
and (ii) an excess cross-power search targeting a well-localized
electromagnetic counterpart,
which does not assume any phase model at all.

The key factors governing detectability are summarized in Table \ref{tab:fal1}.
Four scenarios emerge from \S\ref{sec:fal3a}, 
classified according to whether the maximum $\Omega$ 
(achieved at $t\approx t_{\rm pk}$)
is set by magnetocentrifugal equilibrium or gravitational radiation stalling
(C or G, first column),
and whether gravitational wave emission is dominated by a magnetic mountain
or hydrodynamic instabilities
(M or I, second column).
\footnote{
In reality, the signal is the sum of the M and I components, 
but they are presented separately for the convenience of the reader,
because certain search pipelines may be more sensitive to one component.
For example,
the mountain signal may be easier to detect with a matched filter,
because it is arguably cleaner than the instability signal,
which carries the imprint of nonlinear physics like hydrodynamic turbulence.
See also footnote \ref{foot:fal1}.
\label{foot:fal2}
}
Scalings are presented for $h_{\rm max}$ and $T_{\rm obs}$ 
(third and fourth columns) in terms of
the normalized initial magnetic dipole moment,
$\tilde{\mu} = \mu_{\rm i} / (10^{33}\,{\rm G\,cm^3})$,
and the fallback parameter, $\eta$.
The source lifetime generally satisfies $T_{\rm obs}\gg t_{\rm pk}$,
i.e.\ $h_0(t)$ rises faster than it decays.
It is set by
the propeller effect 
($T_{\rm obs} \approx I\Omega/N_{\rm a}$, with $\omega\approx 1.05$)
or the back reaction from mountain gravitational radiation 
($T_{\rm obs} \approx I\Omega/N_{\rm gw}$)
in the scenarios C or G respectively.
The matched filter SNR is calculated from
${\rm SNR}_{\rm mf}^2 \approx
 32 h_{\rm max}^2 T_{\rm obs} / [375 S_h(f=\Omega/\pi)]$
[e.g.\ equation (13) in \citet{vig09a}],
where $S_h(f)$ is the one-sided detector noise power spectral density
at the observing frequency $f$.
We thereby assume
that most of the power is emitted at twice the spin frequency
\citep{jar98} ---
almost certainly an oversimplification for instabilities
and possibly also for a mountain,
if it wobbles in response to vigorous accretion
[see footnote \ref{foot:fal1} and \citet{pay06b}].
In all table entries the source distance is normalized to $1\,{\rm Mpc}$,
and ${\rm SNR}_{\rm mf}$ is quoted in terms of 
$\tilde{S}_h = S_h(f=0.2\,{\rm kHz}) / (10^{-47} \, {\rm Hz^{-1}})$,
with
$\tilde{S}_h = 1.4$, 0.90, and $6.7\times 10^{-3}$
for zero-detuning high-power Advanced LIGO,
neutron-star-inspiral-optimized Advanced LIGO,
and the conventional Einstein Telescope respectively
[\citet{ben10} and references therein].
Interferometer configurations optimized for $f \lesssim 40\,{\rm Hz}$
(e.g.\ black-hole-inspiral-optimized Advanced LIGO,
xylophone Einstein Telescope)
are not considered in this paper,
where we are interested in signals with $f\gtrsim 0.2\,{\rm kHz}$
and hence $S_h(f) \propto f^2$.
The final column quantifies roughly the reduction in detection distance
expected when replacing a matched filter with an excess cross-power search,
based on the results in Table 1 in \citet{pir12}.

\begin{table}
\begin{center}
\begin{tabular}{cccccc}\hline
 Spin limit \tablenotemark{a} & 
  Signal \tablenotemark{b} & 
  $10^{23} D_{\rm Mpc} h_{\rm max}$ \tablenotemark{c} &
  $T_{\rm obs}$ ($10^2\,{\rm s}$) \tablenotemark{d} & 
  $\tilde{S}_h^{1/2} D_{\rm Mpc} {\rm SNR}_{\rm mf}$ \tablenotemark{e} & 
  $D_{\rm mf}/D_{\rm cp}$ \tablenotemark{f} \\ \hline
 C & M & 
  $2.0 \tilde{\mu}^{2/7} \eta^{60/91}$ &
  $8.8 \tilde{\mu}^{-8/7} \eta^{-30/91}$ &
  $3.7 \tilde{\mu}^{4/7} \eta^{15/91}$ &
  $\sim 14$ \\
 C & I & 
  $6.3 \tilde{\mu}^{3/7} \eta^{20/91}$ &
  {\rm as above} &
  $12 \tilde{\mu}^{5/7} \eta^{-25/91}$ &
  {\rm as above} \\
 G & M & 
  $6.7 \tilde{\mu}^{2/5} \eta^{4/13}$ &
  $1.2 \tilde{\mu}^{-4/5} \eta^{-8/3}$ &
  $3.8 \tilde{\mu}^{4/5} \eta^{-46/39}$ &
  $\sim 6$ \\
 G & I & 
  $7.3 \tilde{\mu}^{2/5} \eta^{4/13}$ &
  {\rm as above} &
  $4.2 \tilde{\mu}^{4/5} \eta^{-46/39}$ &
  {\rm as above} \\
\hline
\end{tabular}
\end{center}
\caption{
Gravitational wave detection: accretion and emission scenarios
}
\tablenotetext{a}{Accretion mechanism that determines the maximum value of $\Omega$
 at $t\approx t_{\rm pk}$:
 magnetocentrifugal equilibrium (C) or gravitational radiation stalling (G).}
\tablenotetext{b}{Gravitational wave emission mechanism:
 magnetic mountain (M) or hydrodynamic instabilities (I).}
\tablenotetext{c}{Peak wave strain $h_{\rm max}$ at $t\approx t_{\rm pk}$;
 $D_{\rm Mpc}$ denotes the source distance measured in ${\rm Mpc}$.}
\tablenotetext{d}{Signal duration $T_{\rm obs}=I\Omega/N_{\rm a}$ (C)
 or $I\Omega/N_{\rm gw}$ (G) in units of $10^2\,{\rm s}$;
 see \S\ref{sec:fal3b}.}
\tablenotetext{e}{Signal-to-noise ratio ${\rm SNR}_{\rm mf}$ 
 for a matched filter search;
 $\tilde{S}_h = S_h(0.2\,{\rm kHz}) / (10^{-47}\,{\rm Hz^{-1}}) $
 denotes the normalized detector noise power spectral density at $0.2\,{\rm kHz}$.}
\tablenotetext{f}{Matched filter ($D_{\rm mf}$) and excess cross-power ($D_{\rm cp}$)
 detection distances;
 ratio calibrated against Table 1 in \citet{pir12}.}
\tablecomments{
In all entries,
$\tilde{\mu} = \mu_{\rm i} / ( 10^{33} \, {\rm G\,cm^3} )$ denotes the
normalized initial magnetic dipole moment,
and $\eta$ is the fallback accretion parameter defined in \S\ref{sec:fal2a}.
}
\label{tab:fal1}
\end{table}

Table \ref{tab:fal1} demonstrates that the prospects for detecting
a nearby protomagnetar remain respectable,
as stated by previous authors
\citep{pir11,pir12},
when the physics of magnetic funnelling and burial is included.
For example, a Local Group object with 
$\mu_{\rm i}=5\times 10^{33}\,{\rm G\,cm^3}$ and $D=4\,{\rm Mpc}$ 
reaches ${\rm SNR}_{\rm mf}=42$ 
in a matched filter search with the Einstein Telescope 
just from its magnetic mountain emission, even before adding instabilities. 
\footnote{
A mountain is detected with false alarm and dismissal
probabilities of 1\% and 10\% respectively, when
$h_0$ exceeds the threshold $11.4 [S_h(f)/T_{\rm obs}]^{1/2}$
\citep{jar98},
corresponding to ${\rm SNR}_{\rm mf}=3.3$.
\label{foot:fal3}
}
The SNR drops to $3.6$ for Advanced LIGO,
still marginally detectable,
while the effective detection distance is $\sim 6$ times smaller
for a more realistic excess cross-power search (see below).
Table \ref{tab:fal1} contains two main trends:
(i)
gravitational radiation stalling leads to higher $h_{\rm max}$
and shorter $T_{\rm obs}$ than magnetocentrifugal equilibrium,
so both evolutionary pathways produce similar SNRs;
and (ii)
the mountain and instability emission are comparable,
with the latter typically being stronger and favored by
higher $\mu_{\rm i}$ and lower $\eta$.

The excess cross-power algorithm
\citep{thr11,pre12},
a generalized form of the stochastic radiometer statistic
\citep{bal06},
does not assume a priori how the phase evolves.
It is therefore a fairer guide to search performance.
It is implemented in two steps:
one computes a spectrogram ${\rm SNR}(t;f)$ of the 
signal-to-noise ratio, which is proportional to the cross-correlation
of the strains at two interferometers (or a larger network more generally);
then one scans the spectrogram for a contiguous track of positive-valued pixels
using a clustering algorithm
(after excising environmental noise artefacts),
in order to get a total ${\rm SNR}$ for the track.
A detection threshold of ${\rm SNR}\approx 23$ can be achieved realistically
with Advanced LIGO,
assuming a false alarm rate of $0.1\%$,
a false dismissal rate of $50\%$,
and a $1\,{\rm ks}\times 1.7\,{\rm kHz}$ on-source region 
divided into $0.5\,{\rm s} \times 1\,{\rm Hz}$ pixels. 
Tripling the on-source time increases the threshold by $\approx 10\%$.

Detailed Monte-Carlo simulations of the excess cross-power SNR
lie outside the scope of this paper;
there is considerable uncertainty surrounding the waveforms,
when hydrodynamic instabilities operate,
and even the magnetic mountain is unlikely to be a static quadrupole,
when $\dot{M}_{\rm a}$ is so high during fallback.
Instead, we calibrate against the results
presented by \citet{pir12}.
Four of the cases presented in Table 1 in the latter reference,
with $\eta =1$ and maximum stellar mass 2.5--$2.9M_\odot$,
are representative of the results in \S\ref{sec:fal3a} and Figure \ref{sec:fal2}:
the emission frequency ranges from $0.79\,{\rm kHz}$ to $2.3\,{\rm kHz}$,
and the duration of the burst ranges from $0.21\,{\rm ks}$ to $0.35\,{\rm ks}$,
consistent with the $\Omega(t)$ evolution underlying Figure \ref{fig:fal2}.
In the four cases, the ratio of detection distances for the matched filter
and cross-power searches ranges from 6.2 to 14.
Other examples computed by \citet{pir12} are consistent with this range;
at one extreme, for $\eta=0.3$, the duration increases to $0.64\,{\rm ks}$
and the detection distance ratio is 19.
As a rough guide, we adopt these results here
(final column, Table \ref{tab:fal1}),
by applying a 14-fold reduction to the detection distance
for magnetocentrifugal equilibrium,
where $T_{\rm obs}$ is longer,
and a six-fold reduction for gravitational stalling,
where $T_{\rm obs}$ is shorter.
When the phase evolution is unknown,
and a template search is prohibitive computationally,
excess cross-power performs better relative to a matched filter
the longer the signal lasts, 
as the $t$-$f$ pixel tracker asserts its advantage.

\section{Black hole formation
 \label{sec:fal4}}
The gravitational wave signals discussed in \S\ref{sec:fal3}
are predicted to truncate suddenly at $t\gtrsim t_{\rm pk}$,
once enough material falls back onto the neutron star to form
a black hole. 
The maximum gravitational mass of a stable protomagnetar,
$M_{\rm max}$,
is given by the nonrotating Tolman-Oppenheimer-Volkoff mass,
$M_{\rm TOV}$,
corrected for centrifugal support from differential rotation.
To a first approximation, one can write
$M_{\rm max} = M_{\rm TOV} ( 1 + \gamma \tilde{\Omega}^2)$,
where $\tilde{\Omega}$ is the angular velocity normalized by
its centrifugal limit,
$\gamma$ is a constant of order unity
\citep{lyf03},
and the nonrotating maximum mass spans the range
$2\lesssim M_{\rm TOV}/M_\odot \lesssim 3$ 
for a selection of popular equations of state allowed by observations
\citep{las14}.

\begin{figure}
\begin{center}
\includegraphics[width=10cm,angle=0]{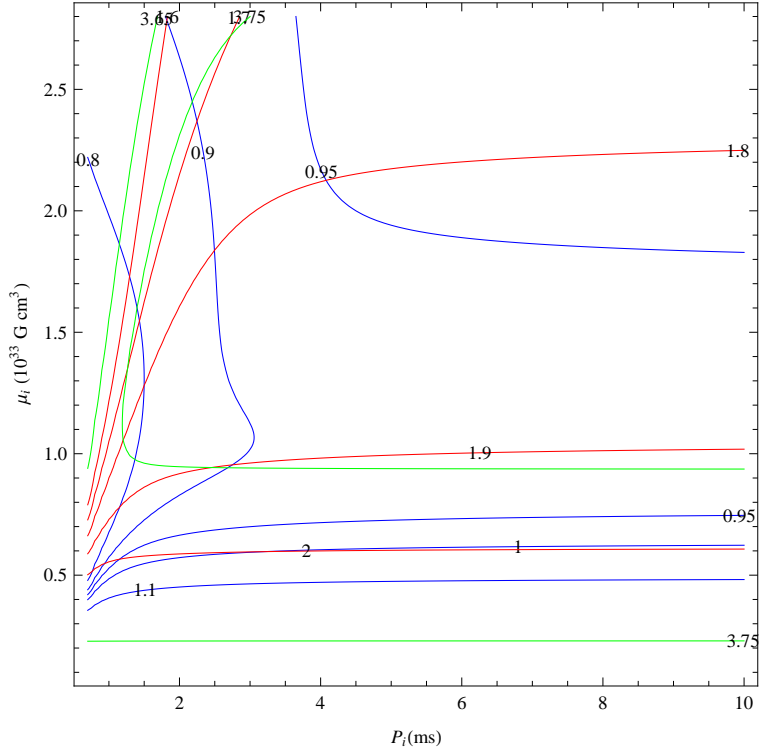}
\end{center}
\caption{
Contours of nominal accreted mass, $M(10^4\,{\rm s})-M(0)$,
in solar masses,
as a function of initial spin period $P_{\rm i}$ 
(in units of ${\rm ms}$) and
initial magnetic dipole moment $\mu_{\rm i}$
(in units of $10^{33}\,{\rm G\,cm^3}$),
for
$\eta=0.3$ (blue),
$1.0$ (red),
and
$3.0$ (green).
All curves include mountain and instability contributions to $N_{\rm a}$
and $N_{\rm gw}$ and exclude $\mu$ reduction by burial;
if $\mu$ is buried,
there is no propeller effect, and one obtains  
$M(10^4\,{\rm s})-M(0) =
 \int_0^{10^4\,{\rm s}} dt' \dot{M}_{\rm a}(t')
= 2.0 M_\odot$ $3.0 M_\odot$, and $4.3 M_\odot$ for
$\eta=0.3$, $1.0$, and $3.0$ respectively.
Parameters:
$\beta_{\rm c}=0.14$,
$M_{\rm c} = 3\times 10^{-3} 
 ( \mu_{\rm i} / 10^{33}\,{\rm G\,cm^3})^2$,
$M(0)=1.4M_\odot$, $R=10\,{\rm km}$.
}
\label{fig:fal3}
\end{figure}

Figure \ref{fig:fal3} shows how the nominal mass accreted onto the
neutron star varies with initial magnetization and spin,
before taking black hole formation into account.
Contours are drawn for three fallback scenarios:
$\eta=0.3$ (blue), $1.0$ (red), and $3.0$ (green).
The results are to be compared with Figures 3 and 4 in \citet{pir11}.
We include mountain and instability contributions to the torques
but not $\mu$ reduction by magnetic burial,
as for the solid contours in the bottom panel of Figure \ref{fig:fal2}.
If $\mu$ is buried,
the propeller effect is suppressed for all $P_i$ and $\mu_{\rm i}$
across the plotted region;
all the infalling matter,
$\int dt'\, \dot{M}_a(t')$, finds its way onto the protomagnetar,
which turns into a black hole at $t\sim t_{\rm pk}$
for all $\eta \geq 0.3$.
If $\mu$ is not buried, the formation of a magnetic mountain
still modifies the results of \citet{pir11} to some extent
through $N_{\rm gw}$.
\footnote{
This is an unlikely scenario prima facie, 
because the growth of $\epsilon$ goes hand in hand with $\mu$ reduction
in the rigorous theory of magnetic burial developed for low-mass
X-ray binaries
\citep{pay04,vig09a,pri11}.
Nevertheless we mention it for completeness,
in case $\epsilon$ and $\mu$ decouple for some
reason during supernova fallback in the high-$\dot{M}_{\rm a}$ regime.
}

In general, the trends in Figure \ref{fig:fal3} are qualitatively similar
to those observed in Figures 3 and 4 in \citet{pir11}. 
The main difference is that the nominal accreted mass is 
$\approx 20\%$ higher at $\mu_{\rm i}=1\times 10^{33}\,{\rm G\,cm^3}$,
and
$\approx 50\%$ higher at $\mu_{\rm i}=3\times 10^{33}\,{\rm G\,cm^3}$,
because the mountain gravitational wave torque,
$N_{\rm gw} \propto \mu_{\rm i}^4$,
decelerates the star and moderates the propeller effect.
When fallback is vigorous, e.g.\ $\eta = 3$, the nominal accreted mass
exceeds $3.5M_\odot$ over the plotted region,
and a black hole always forms.
When fallback is less vigorous,
e.g.\ $\eta = 0.3$,
a black hole still forms in most parts of the figure except the top left.
Taking $M_{\rm max}=2.2M_\odot$, 
at the lower end of the stable range,
black hole formation is prevented for
$\mu_{\rm i} \gtrsim 0.8\times 10^{33}\,{\rm G\,cm^3}$ 
and 
$P_{\rm i} \lesssim 2\,{\rm ms}$ 
for $\eta=0.3$.
Taking $M_{\rm max}=3.5M_\odot$, at the upper end of the stable range
after correcting for differential rotation
\citep{las14},
black hole formation is prevented across the whole plotted region
for $\eta=0.3$ and for
$\mu_{\rm i} \gtrsim 0.5\times 10^{33}\,{\rm G\,cm^3}$ 
for $\eta=1.0$.

Recent observations of a sharp truncation in the X-ray flux 
from some short gamma-ray bursts,
$\sim 10^2\,{\rm s}$ after the initial trigger,
have been interpreted as evidence that we are seeing 
a protomagnetar forming in a binary neutron star coalescence event 
and collapsing to form black hole
\citep{row13}.
Although this phenomenon is not described by the analysis in this paper,
which is motivated by core-collapse supernovae, it is related in two ways.
First,
\citet{las14} have proposed a multimessenger experiment 
targeting short gamma-ray bursts,
in which the X-ray light curve and
an Advanced LIGO measurement of the progenitor chirp mass
are combined with a simple magnetic braking model
and the known mass distribution in neutron star binaries
to constrain the nuclear equation of state.
It is worth investigating whether a similar experiment is feasible
in the supernova context, 
as suggested originally in {\S}5 of the paper by \citet{pir12}.
Second,
in the gamma-ray burst context,
it may be worthwhile to generalize the magnetic braking model 
assumed by \citet{las14} to include
some of the ingredients discussed in this paper,
such as hydrodynamic instabilities,
the fallback accretion torque (including in the propeller regime),
the gravitational wave torque,
magnetic funnelling,
magnetic mountain growth, 
and magnetic dipole moment reduction by burial.

\section{Conclusion
 \label{sec:fal5}}
In this paper we revisit the scenario of protomagnetar spin up
by supernova fallback studied by
\citet{pir11} and \citet{pir12},
in which hydrodynamic instabilities radiate gravitational waves,
and the magnetocentrifugal propeller effect impedes black hole formation.
We add one extra ingredient:
magnetic funnelling of the accretion flow.
Following closely the analysis in the original papers,
we show that magnetic funnelling modifies the scenario in two ways.
First,
a polar magnetic mountain forms,
confined by the accretion-compressed equatorial magnetic field
as in low-mass X-ray binaries
\citep{pay04},
whose mass quadrupole moment is substantial,
with $\epsilon = 3\times 10^{-3} (\mu_{\rm i}/10^{33}\,{\rm G\,cm^3})^2$.
The associated, quasimonochromatic gravitational wave signal
is somewhat weaker than the instability signal at its peak
but can last somewhat longer (see Table \ref{tab:fal1}),
contributing comparably to the SNR and
boosting the likelihood of detection.
Figure \ref{fig:fal2} and Table \ref{tab:fal1} present
the wave strain and SNR as functions of $P_{\rm i}$ and $\mu_{\rm i}$.
They show that the peak wave strain and burst duration are chiefly
controlled by whether the initial, fallback-driven spin up
of the protomagnetar stalls in response to
the magnetocentrifugal or gravitational wave torque,
with the latter dominating 
for $\mu_{\rm i} \gtrsim 1\times 10^{33}\,{\rm G\,cm^3}$.
Second,
as in low-mass X-ray binaries,
growth of a polar magnetic mountain is accompanied by
reduction of the stellar magnetic dipole moment,
even as the local magnetic field at the equator intensifies
\citep{pay04}.
As $\mu$ diminishes, the propeller effect rapidly shuts off,
and most of the supernova debris falls back onto the protomagnetar,
assisting black hole formation.
Figure \ref{fig:fal3} presents the nominal accreted mass
as a function of $P_{\rm i}$, $\mu_{\rm i}$, and $\eta$.
Gravitational wave emission is dominated by the magnetic mountain
for $\mu_{\rm i} \gtrsim 1\times 10^{33}\,{\rm G\,cm^3}$
and instabilities
for $\mu_{\rm i} \lesssim 1\times 10^{33}\,{\rm G\,cm^3}$.

We emphasize in closing that the results
in Figures \ref{fig:fal2} and \ref{fig:fal3} and Table \ref{tab:fal1}
are indicative only.
The model is idealized.
The equations of motion for $\Omega(t)$ and $M(t)$ implicitly
assume magnetized, thin-disk accretion,
which may not hold at the high accretion rates characteristic
of supernova fallback.
For example,
the magnetic propeller may work less efficiently in a messy,
high-$\dot{M}_{\rm a}$ accretion environment,
whether or not $\mu$ is reduced by burial.
The degree to which hydrodynamic instabilities self-adjust 
to cancel $N_{\rm a}$ is a subtle issue;
further study is needed before the approximation made here,
namely perfect cancellation for $\beta>\beta_{\rm c}$,
can be accepted as valid 
(see footnote \ref{foot:fal1}).
The complicated waveforms produced at high $\dot{M}_{\rm a}$
by hydrodynamic instabilities and a wobbling magnetic mountain
\citep{pay06b}
are unknown at the time of writing;
they are approximated by toy waveforms 
(or circumvented by computing the total radiated power) 
in this paper and by previous authors.
Precession is neglected,
even though it is likely to modify both the waveform
\citep{pay06b} and the torques (especially $N_{\rm a}$),
when $\dot{M}_{\rm a}$ is high.
In these and other respects, therefore, the model is emphatically not
a substitute for large-scale, magnetohydrodynamic simulations.
Its value rests in drawing attention to the physical implications 
of a single, new effect
--- magnetic funnelling ---
by comparing directly with otherwise identical models computed previously
\citep{pir11,pir12}.
In summary, when magnetic funnelling is included,
(i)
the gravitational wave amplitude and duration are multiplied by 
factors of roughly $0.3$--$0.9$ and $1$--$7$ respectively,
boosting the SNR by up to $\sim 3$ times overall,
and (ii)
a black hole is more likely to form,
because the propeller phase is suppressed,
and $\gtrsim 50\%$ of the infalling matter lands on the compact object.

\acknowledgments
AM thanks the Theoretical Astrophysics Group at the 
Eberhard Karls Universit\"{a}t T\"{u}bingen
for its generous hospitality while the paper was started.
This work was supported by an
Australian Research Council Discovery Project grant
and a Group of Eight Germany Joint Research Cooperation grant.

\bibliographystyle{mn2e}
\bibliography{supernova}

\end{document}